\documentclass[useAMS,usenatbib]{mn2e}

\usepackage{graphicx}

\newcommand{\simgt}{\lower.5ex\hbox{$\;\buildrel>\over\sim\;$}}
\newcommand{\simlt}{\lower.5ex\hbox{$\;\buildrel<\over\sim\;$}}

\newcommand{\mstar}{\ensuremath{M_{\rm star}}}

\newcommand{\ha}{H$\alpha$}
\newcommand{\hb}{H$\beta$}

\newcommand{\aap}{A\&A}
\newcommand{\apj} {ApJ}

\newcommand{\aj} {AJ}
\newcommand{\araa} {ARA\&A}
\newcommand{\mnras}{MNRAS}

\newcommand{\pasp}{PASP}

\begin{document}
\title[The FMR in the cluster 7C 1756+6520]{LUCIFER@LBT view of star-forming galaxies in the cluster 7C 1756+6520 at z$\sim$1.4}
\author[L. Magrini et al.]
  {
Laura Magrini$^1$, Veronica Sommariva$^1$, Giovanni Cresci$^1$, Eleonora Sani$^1$,\newauthor 
 Audrey Galametz$^2$,  Filippo Mannucci$^1$, Vasiliki Petropoulou$^3$, Marco Fumana$^4$
\\
  $^{1}$ INAF - Osservatorio Astrofisico di Arcetri, Largo E. Fermi 5, I-50125, Firenze, Italy\\
  $^2$ INAF Osservatorio Astronomico di Roma, Via Frascati 33, I-00040, Monteporzio (Roma), Italy\\
  $^3$ Instituto de Astrofisica de Andalucia-C.S.I.C., Glorieta de la Astronomia, E-18008 Granada, Spain\\ 
  $^4$ INAF - Istituto di Astrofisica Spaziale e Fisica cosmica, Milano, Italy
}

 

\date{Accepted ?. Received ?; in original form ?}

\pagerange{\pageref{firstpage}--\pageref{lastpage}} \pubyear{2012}

\maketitle

\label{firstpage}

\begin{abstract}
Galaxy clusters are key places to study the contribution of {\it nature} 
(i.e.~mass, morphology) and {\it nurture} (i.e.~environment) in the formation 
and evolution of galaxies.
Recently, a number of clusters at z$>$1, i.e. corresponding to the
first epochs of the cluster formation,   has been discovered and confirmed spectroscopically.
We present new observations obtained with the {\sc LUCIFER} spectrograph at Large Binocular Telescope (LBT) 
of a sample of star-forming galaxies associated with a large scale structure around the radio galaxy 7C1756+6520 at z=1.42. 
Combining our spectroscopic data and the literature photometric data, we derived some of the  properties of these galaxies: star formation rate,  metallicity and stellar mass.  
With the aim of analyzing the effect of the cluster environment on galaxy evolution, 
we have located the galaxies  in the plane of the so-called Fundamental Metallically Relation (FMR), which is known not to evolve
 with redshift  up to z$=2.5$ for field galaxies, but it is still unexplored in rich environments at low and high redshift.
We found that the properties of the galaxies in  the cluster 7C 1756+6520 are compatible 
with the FMR which suggests that the effect of the environment on galaxy metallicity 
at this early epoch of cluster formation is marginal.
As a side study, we also report  the spectroscopic analysis of a bright AGN, belonging to the cluster, which shows a significant outflow of gas. 

\end{abstract}

\section[]{Introduction}

Located at nodes of the cosmic web, clusters of galaxies are the largest collapsed structures in
the Universe with total masses up to 10$^{15}$ M$_{\odot}$. Over 80\% of their mass resides in the form of dark
matter, whereas  the remaining mass is composed of baryons, most of which (about 85\%) is a diffuse, hot T
$>$ 10$^{7}$ K plasma, the intracluster medium (ICM) \citep[e.g.,][]{arnaud09}. 
It is known that galaxy evolution is closely linked to the formation of structures in the  Universe. 
In the Local Universe, both galaxy morphology and star formation history have been found to correlate with the environment, with passive 
early-type galaxies dominating the dense regions \citep[e.g.,][]{dressler80,finn05}. 
This points towards a differential evolution between isolated and cluster galaxies \citep[e.g.,][]{poggianti06,bolzonella10}
which results in a strong environmental dependence of galaxy properties  in the local Universe \citep[e.g.,][]{diserego05,chung09,petro12}.

In this framework, many questions have risen during the last decades about the effect of the 
environment in galaxy evolution, for example:  when were the visible parts of
galaxies assembled, when were the stars formed, and how
did this depend on the environment? Is it {\em nature} (i.e. mass, morphology) and/or {\em nurture}
(i.e. environment)  that governs the evolution of a galaxy? which is the origin of the morphological segregation? 
By studying different stages of their
evolution, galaxy clusters are among the best tools to seek answers to  these questions.

Most of the recent studies have been devoted to clusters at z$<1$. From these studies it has been found  that 
the environmental dependence of galaxy properties is already strong at z$=1$, with  red galaxies being 
the dominant population \citep[e.g.,][]{blakeslee03,nakata05,Lidman08}. 
Due to the smaller number of galaxy clusters confirmed
at z$>$1, it has been challenging until now to reproduce these studies at higher 
redshift. However, the number of spectroscopically confirmed clusters at $z > 1$ 
has significantly increased in the last few years thanks to new efficient selection
techniques e.g.~wide surveys in the mid-infrared \citep{eisenhardt08,wilson09,papovich08}, 
deep X-rays observations of the ICM \citep{Lidman08,hilton09} or targeted studies
in the field of radio galaxies \citep[][and references therein]{g09,g10b}.
This last selection technique has permitted the discovery of a galaxy cluster
associated with 7C1756+6520 --- a radio galaxy at z=1.42 --- on which this paper
is focussed (see Section~2). To date, the two highest redshift (spectroscopically confirmed)
galaxy clusters, discovered using mid-infrared selection techniques, were found at 
z=1.62 \citep{papovich10,tanaka10} and z=2 \citep{gobat11}. The 
increasing number of proto-clusters (associated with high-redshift radio galaxies)
at z$>$2 and up to z$\sim$6 \citep[among others,][]{venemans07,hatch11,kuiper11} opens even broader perspectives for the study of galaxy properties
dependance with environment and cosmic time.


The metal content of galaxies is an important tracer of their evolutionary stage. 
In the Local Universe,  it is now well established  \citep{lequeux79} that gas-phase metallicity and stellar mass 
of galaxies are tightly related in a mass-metallicity relation, with  more massive galaxies being  also more metal rich
\citep[see, for example,][]{tremonti04}. 
Several studies have demonstrated that this relation evolves  with
redshift, with metallicity decreasing with redshift at a given mass \citep[e.g.,][]{erb06,maiolino08,mannucci09}. A similar relation 
was observed also between stellar metallicities and stellar masses \citep{sommariva12}.
However, a more fundamental relation seems to be in place: \citet{mannucci10} found that metallicity depends not only
on mass, but also on star formation rate (SFR), with more star-forming galaxies, showing lower metallicities at 
a given stellar mass. 
At z$=$0, local  galaxies define a surface in the mass-SFR-metallicity space \citep[see Fig.~2 in][]{mannucci10}, 
called the  Fundamental Metallicity Relation (FMR), in which the spread of the metallicity around this relation is lower that 
in the {\em classical} Mass-Metallicity (Z) Relation (MZR) (0.05~dex vs 0.10~dex). 
In contrast to the MZR, {\em the FMR does
not evolve with redshift}, i.e., all the observed galaxies up to z$=2.5$ are found to follow the
same local relation \citep[see, e.g.,][for an extension towards low masses]{cresci12,mannucci11}.
The origin of the FMR is not still completely clear, but it is probably due to a
combination of several effects: infall of pristine gas, outflow of enriched material, different
efficiencies  of star formation for different masses ({\em chemical downsizing}). 
Its existence and the lack of evolution up to z$=2.5$, $\sim$80\% of the Hubble time, means that this is really a fundamental
relation tightly linked to the {\em nature} of the processes of galaxy formation \citep[see, e.g.,][]{dave11,campisi11,dayal12}.

The aim of the present work is to derive, for the first time, gas-phase metallicities, stellar masses, and SFRs 
of a sample of star-forming galaxies belonging to one of the farther 
spectroscopically confirmed clusters of galaxies, at z$\sim$1.4.
This will allow us to study, for the first time, the FMR in a rich environment at high redshift
analyzing if the cluster membership alters the location of galaxies along the FMR plane. 
 In particular, different effects on the FMR in dense environments  are expected if the dominant 
processes are the infall or  the outflow:  for example, since the ICM is metal rich,  
an infall of already enriched gas could have an effect on the FMR radically different than an infall of pristine gas.

In addition, the mask design has allowed us to include the observations of an active galactic nucleus (AGN( belonging to the cluster, for which we could  
analyze the  line profiles identifying the origin of the  
blue wings in its emission lines. 
The paper is structured as follows: 
in Sec.\ref{sec1} we present the main characteristics of the cluster  7C 1756+6520 and its literature data. 
In Sec.\ref{sec_targets}  we describe the observations and the selection of the targets to be observed spectroscopically, and in Sec.\ref{sec_obs} the spectroscopic 
observations and how the parameters of galaxies (stellar mass, SFR, and metallicity) are derived (Sec.\ref{sec_par}). In Sec.\ref{sec_fmr} we discuss the FMR for the cluster galaxies, and in Sec.\ref{sec_agn}
we present the spectroscopic analysis of an AGN belonging to the cluster. 
Finally, in Sec.\ref{sec_discu} we give our conclusions. 

\section[]{A cluster at z$\sim$1.42 around 7C 1756+6520}
\label{sec1}

\citet{g09} (hereafter, G09)  made
use of the well known BzK selection technique \citep{daddi04} to isolate z$>$1.4 galaxy candidates 
in the surroundings of 7C 1756+6520, a radio galaxy at z$=$1.42 and found an overdensity of z$~$1.4
candidates in this field.
The existence of this  large scale structure of galaxies was then confirmed by \citet{g10} (hereafter, G10) through  optical Keck/DEIMOS spectroscopy. 
They  assigned a redshift of z$=1.4156$  to the radio galaxy, and about twenty galaxies with spectroscopic redshifts consistent
with the redshift of 7C 1756+6520 were confirmed. 
They found that seven of these galaxies have
velocity offsets v$<$1000 km s$^{-1}$ relative to the redshift of
the radio galaxy and are within $\sim$2~Mpc from the radio galaxy, forming a first sub-structure located around the radio galaxy. 
In addition they found a second sub-structure, located at the East  of the radio galaxy, with a redshift of z$\sim$1.44.

\section[]{Observations}
\label{sec_targets}
The observations presented here were carried out with the Large Binocular Telescope (LBT), located on Mount Graham, Arizona (Hill et al. 2006), 
using the spectrograph LUCIFER \citep{ageorges10,seifert10}.
The NIR spectrograph LUCIFER is   mounted on the bent Gregorian focus of the left primary mirror with a  wavelength coverage  from 0.85 to 2.4 
$\mu$m (zJHK bands) in imaging, long-slit, and multiobject spectroscopy modes.

\begin{figure*}
\begin{center} 
\includegraphics[width=15.2truecm]{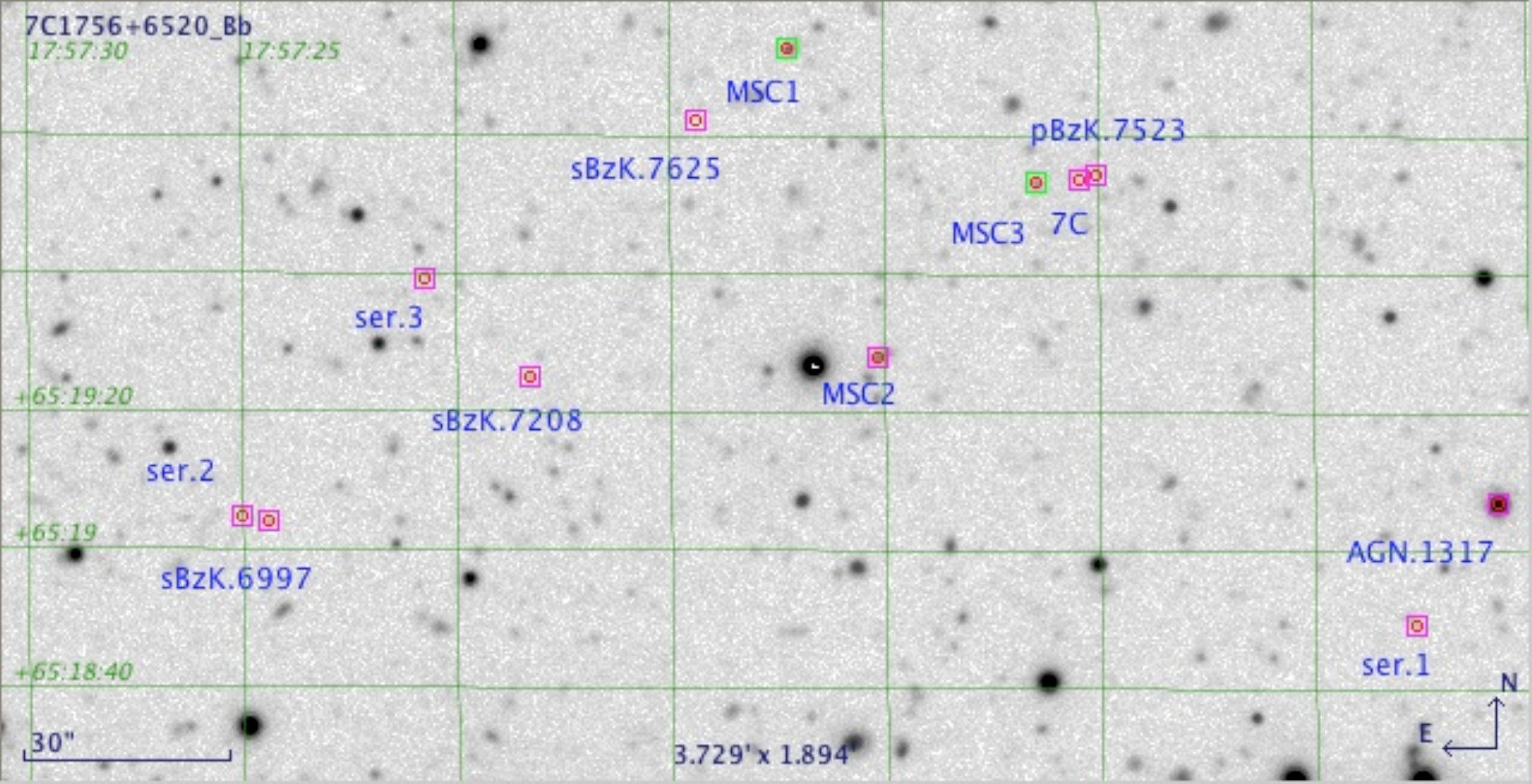} 
\caption{The location of the observed galaxies in the B-band image. North in on the top, East to the left. The field of view is $\sim$3.7'$\times$ $\sim$1.9'.  }
\end{center}
\label{fig_1}
\end{figure*}

The mask design was done starting with the list of star-forming galaxies 
and with the B-band image of G10. 
To maximize the number of galaxies to  be observed with a single LUCIFER mask, 
we centered the mask  at R.A. 17:57:13.03, Dec. 65:19:13.5 (J2000.0), with a rotation angle respect to the North of 250 degrees. 
We selected seven star-forming galaxies, identified by G10 from the optical 
spectroscopy thanks to their intense [OII] 3727 \AA\ emission line, 
and two AGNs, already classified by G10 thanks to several broad band features. 
We also allocated other slits to galaxies not previously studied, for which we obtained spectroscopic redshifts. 
In Fig.\ref{fig_1}  we show with squares  the location of the galaxies that were observed spectroscopically.  In Table~\ref{tab_target} we report the identification name for each galaxy (from G10 when available, otherwise they are identified with the name {\sc MSC} followed by a number), 
the J2000.0 R.A. and Dec. coordinates, and the spectroscopic redshifts, from literature (G10) and from the present work, when available. 
\begin{table*}
\caption{Observed targets:  Col. 1 object name, Cols. 2, 3, coordinates(J2000), Col. 4 spectroscopic redshifts 
from Galametz et al 2010, Col.5  spectroscopic redshifts from this work. We assigned a quality flag A or B to the measured
redshifts.  }
\centering
\begin{tabular}{llllll}
\hline
\multicolumn{5}{c}{}\\
 id  &  R.A. J2000.0              & Dec. J2000.0           &  redshift  (G10)   & redshift (this work)       \\
\multicolumn{5}{c}{}\\
\hline
serendip.1 		&17:56:57.67 	&+65:18:49.45 & 1.4150$\pm$0.0005 & 1.453 $\pm$0.001 (A)\\
serendip.2 		&17:57:25.00 	&+65:19:04.83 & 1.4157$\pm$0.0010 & 1.416 (B)\\
sBzK.6997 		&17:57:24.43 	&+65:19:03.87 & 1.4157$\pm$0.0006 & 1.416 (B)\\
pBzK.7523$^a$ 	&17:57:05.04 	&+65:19:54.50 & 1.4244$\pm$0.0004 &-\\
sBzK.7625$^a$  	&17:57:14.41 	&+65:20:02.40 & 1.4366$\pm$0.0001 &-\\
sBzK.7208 		&17:57:18.31 	&+65:19:24.94 & 1.4374$\pm$0.0002 &1.4376$\pm$0.0005 (A) \\
serendip.3 		&17:57:20.76 	&+65:19:39.14 & 1.4379$\pm$0.0007 &1.4371$\pm$0.0005 (A) \\
AGN.1206$^a$ 	&17:57:13.08	&+65:19:08.37 & 1.4371$\pm$0.0002 &-\\
AGN.1317		&17:56:55.75 	&+65:19:07.00 & 1.4162$\pm$0.0005 &	1.4168$\pm$0.0005	(A)		\\
MSC2 			&17:57:10.16	&+65:19:28.09 & -		                      & 1.4556$\pm$0.0005 (A) 	\\
\hline
MSC1 			&17:57:12.23	&+65:20:12.68 & -                                    & 1.551$\pm$0.001 (A)			\\
MSC3 			&17:57:06.46	&+65:19:53.29 & -                                    & 2.372$\pm$0.001 (A)			\\	
\multicolumn{5}{c}{}\\
\hline
(a) no detected emission lines \\
\end{tabular}
\label{tab_target}
\end{table*}

\subsection[]{Spectroscopic observations}
\label{sec_obs}
We used  MOS mode, 
 obtaining spectroscopy of  ten star-forming galaxies in the two main groups of the cluster,  
(i.e. the galaxy cluster centered on the radio galaxy at redshift z=1.4156, and a compact galaxy sub-group at $z\sim1.437$, see also Sec.~\ref{sec1}):  seven spectroscopically confirmed by G10 and three without previous observations; 
we obtained spectra also for two AGNs. 
The spectroscopic observations were taken in the H band  with slit-width of 1 arcsec aiming to include several metallicity diagnostic lines:  H$\alpha$, the nitrogen 
doublet {\sc[NII]}, and the sulphur doublet {\sc [SII]}
The grating 210-zJHK, with an appropriate tilt,  allowed us to cover a spectral range from  1.48 to 1.67 $\mu$m in H band. 
The tilt was set to have a full spectral coverage in the spectral region of our interest  for all slits. 
The resolving power of 210-zJHK in the H bands is R$=$7838, meaning that the spectral resolution is 
$\sim$4 \AA\,  suitable to separate  emission lines and skylines. 
We adopted the  {\em nodding} procedure, i.e. observing the object of interest  at different positions along the slit. 
This allowed us to do a proper sky subtraction. 
The total exposure was 5.1~hr in the H band during the nights of June 25 and 26, 2011. A total of 61 exposure, each of 300~s, were taken. 
We also obtained observations in the J band during the nights  from June 26 to 28, 2011, with 52 exposure of 300~s each.  
Unfortunately the observations in the J band were obtained under unfavorable 
weather conditions, with sky  partially covered by clouds, and they could not be used for our science purpose. 
The data were spectroscopically calibrated with a standard star observed during the same nights of the science observations. 

\subsection{Data reduction}
Data reduction was done with the Lucifer Spectroscopic Reduction pipeline which is
described for the first time in the present work. More details can be found at {\tt  http://lbt-spectro.iasf-milano.inaf.it/lreducerInfo/}
It consists in a set of tasks, based on VIPGI \citep{scodeggio05} recipes,
working on Linux systems and written in C and Python.

The preliminary steps of the pipeline workflow concern the creation of master calibration frames:
{\em i) } a bad pixel map was created using dark and flat field frames for each observing run; 
{\em ii)} a master dark was produced  combining several dark frames; 
{\em iii)} a master flat was created averaging a set of spectroscopic flats, and it was used to perform the pixel to pixel 
correction on the scientific frames. It also allowed to automatically locate slits on the scientific frame. 
{\em iv)} an inverse dispersion solution was created starting from calibration arc lamps.
{\em v)} a sensitivity function, as well as telluric absorption correction, were obtained using a telluric star observed close in 
time and in airmass to the scientific frames.

\begin{figure}
\begin{center} 
\includegraphics[width=8.5truecm]{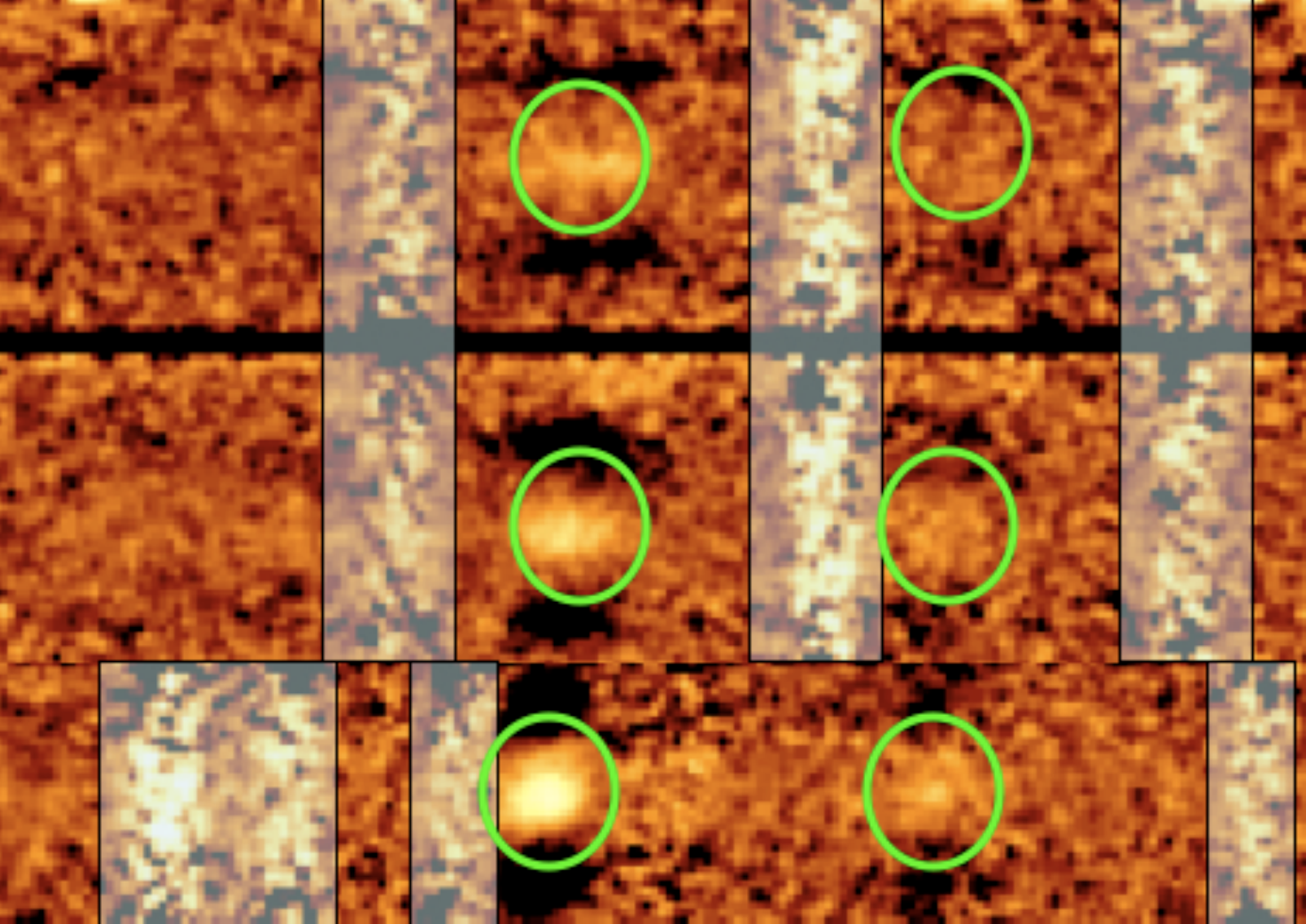} 
\caption{An example of the 2D spectra, namely, from the top, Ser.1, sBzK.7208, and MSC1.  The H$\alpha$ and [NII] are shown within (green) circles. The sky emission lines are (grey) shaded.  
The negative signatures above and below the emission lines are due to the nodding procedure. }
\end{center}
\label{fig_spectra}
\end{figure}

Once all calibration frames were available, cosmic rays and bad pixels were removed, and dark and flat field corrections were applied, 
then the spectra were extracted. In this step slit curvatures were removed, slits were extracted and wavelength calibrated. The wavelength calibration {\em rms} for our frames 
was lower than 0.2 \AA. Sky subtraction, based on the Davies sky subtraction algorithm \citep{davies07}, was done on 2D extracted, wavelength calibrated spectra.
Further sky subtraction residuals were removed subtracting the median computed along the spatial direction for each spectrum column.
Finally, the wavelength and flux calibrated spectra
obtained in the different nights were combined together to obtain a
stack of 2D sky subtracted spectra.
Single spectra were extracted from the 2D stacked frame, using a Horne optimal extraction \citep{horne86}.
An example of the extracted 2D spectra is shown in Fig.\ref{fig_spectra}.

Emission-line fluxes were measured with the task {\sc SPLOT} of {\sc IRAF}. 
We derived the redshifts with  a gaussian fit of the brightest emission-line for each target, which allowed us to find the central wavelength of the line. 
The uncertainties on the spectroscopic redshifts are due to the combination of the uncertainty on the fitting and the  uncertainty on the wavelength calibration 
(0.15 \AA). Following G10, we assigned a quality flag ``A'' or ``B'' to the measured redshifts: ``A'' indicates a secure $z_{\rm spec}$ based on at least two spectral features (H$\alpha$ and [NII])
 ``B'' indicates high-level confidence in the $z_{\rm spec}$ based on a single  spectral feature, usually  H$\alpha$. 
The observed wavelengths in \AA\ and line fluxes in  erg cm$^{-2}$ s$^{-1}$ are shown in Table~\ref{tab_fluxes}, while the $z_{\rm spec}$  are in Table\ref{tab_target}. 

\begin{table}
\caption{Measured emission line fluxes}
\centering
\begin{tabular}{lllll}
\hline
\multicolumn{5}{c}{}\\
Galaxy &   &  wavelength            & flux & FWHM                \\
 &	& (\AA) & (erg/cm$^{2}$s)   & (\AA) \\
\multicolumn{5}{c}{}\\
\hline
\hline
Cluster memb. &&& &\\
serendip.1 & H$\alpha$	& 16101.31	& 6.9e-17	& 21. \\
&{\sc [NII]}	     		& 16149.52	& 1.7e-17	& 5.  \\
\hline
sBzK.7208 	& H$\alpha$     	&     15998.00  &  1.0e-16   & 17. \\
&{\sc [NII]}		  	         &   16052.07     & 1.8e-17   & 5. \\
\hline
serendip.3 	&Ha     	         &15994.92 & 1.2e-16 & 10. \\   
&{\sc [NII]}	  	     		& 16046.28 & 4.0e-17   & 10. \\
\hline		       
MSC2         	& H$\alpha$      & 16116.03 	&   1.8e-17 & 10.\\
&{\sc [NII]}   	      & 16167.00          &$<$6e-18$^a$\\
\hline
serendip.2 	&H$\alpha$	& 15860	& $<$4e-18$^a$ &	 \\
\hline
sBzK.6997 	&H$\alpha$	& 15860	& $<$4e-18$^a$ &	 \\
\hline
AGN.1317 &H$\beta$ b&    11724.85 & non cal.   & 78. \\
&{\sc [OIII]}	  &          11978.53 & non cal.  & 9.  \\
&{\sc [OIII]} b	  &   11968.86 & non cal.  & 24.  \\
&{\sc [OIII]}	  &           12094.50 & non cal.    & 9. \\
&{\sc [OIII]} b	  &   12084.83 & non cal.   & 24. \\
&H$\alpha$      & 15861.04 & 2.8e-16 & 9.\\   
&{\sc [NII]}	   	               & 15812.72 & 5.2e-17 &9.    \\
&{\sc [NII]}	   	               & 15906.94 & 9.3e-17  & 9.   \\
&H$\alpha$ b &    15851.38 & 9.7e-16  & 121.  \\
&{\sc [SII]}	   	           &16228.27  &  4.4e-17  &9.  \\
&{\sc [SII]}	  	     	     &16264.51 & 2.0e-17  &9.   \\
\hline
\hline
non memb. && &\\
\hline
MSC1 	&H$\alpha$ &   	     16743.2 & 1.9e-16   & 8.\\
&{\sc [NII]}	             &   	     16796.2 & 4.9E-17  &10.  \\
\hline
&{\sc [OII]}	  &         12572.77  & non cal.   & 11.\\
MSC3 	&{\sc [OIII]}   &    	     16883.93 & 1.2E-16 & 10.     \\
\multicolumn{4}{c}{}\\
\hline
\end{tabular}
(a) upper limit obtained measuring the continuum flux in the same spectral region where the emission line is expected
\label{tab_fluxes}
\end{table}

\section{Parameters of the galaxies: stellar mass, metallicity, and SFR}
\label{sec_par}
The following sections present the methods we adopt to obtain an 
estimation of the stellar mass, oxygen abundance and star formation 
rate for the studied galaxies. These derived galaxy parameters are 
summarized in Table\ref{tab_fmr}.

\subsection[]{The stellar masses}
\label{mass}

We estimated the stellar masses  by fitting  the multicolour spectral energy distribution (SED), 
obtained with  the observed magnitudes in 5 photometric bands:
Palomar-LFC B and z, CFHT/WIRCAM J and $K_{s}$ ,
and Spitzer IRAC at 3.6 $\mu$m, 4.5 $\mu$m.
Details of photometric catalogs and data reduction are given in \citet{g09}.   
For a straightforward comparison with the previous studies
on the FMR, we computed the masses with the same 
procedure described in \citet{cresci12}.
The stellar masses were obtained using the code HyperZmass 
\citep{pozzetti07,pozzetti10} a modified version of the {\sc HyperZ} code \citep{bolzonella00}
based on the SED  fitting technique:  for a known spectroscopic redshift 
we compute the best fitting SED between observed and model 
fluxes by using the $\chi^{2}$ minimization.
Along with the best fitting SED and its normalization, HyperZmass
provides an estimate of the star formation rate, extinction, stellar population age and
stellar masses contained in each galaxy,  for further  details see \citet{cresci12}.
We adopted the \citet{bruzual03} code for spectral synthesis
models,  using its low resolution
version with the Padova 1994 tracks. 
The models were produced assuming a  Chabrier Initial Mass Function (IMF) 
\citep{chabrier03} with an upper mass limit of 100$M_{\odot}$.
We have adopted  the extinction law of \citet{calzetti00}  with $A_{V}$ 
varying between 0$-$4 mag, and fixing the redshift to the spectroscopic value.
We used smoothly and exponentially decreasing Star Formation Histories
(SFHs) with time scale $\tau=[0.1,\infty]$ and age $t=[0.1,20]$ in Gyr.
The stellar masses are reported in Table~3 and an example of our fitting is shown in Fig.~2.
The error on the stellar mass is obtained from the SED fitting, while the uncertainty  on $A_{V}$ is 
the formal error due to the step in extinction adopted by {\sc HyperZ}. 
The latter errors are indeed lower limit on the uncertainty on the extinction.  
The major source of error is due to adopt  a value of  $A_V$ obtained from SED fitting, thus 
valid for the stellar component, also to the gaseous component. 
It is known that dust extinction for nebular lines could be  larger by a factor of $\sim$2 \citep[e.g.,][]{calzetti00}.  
Although this effect is still uncertain at high redshift,
recently \citet{forster09} suggest that the H$\alpha$ based
SFRs, with extra dust attenuation by a factor of $\sim$2, are in good
agreement with those derived from the SED fitting for their {\sc SINS}
galaxies at z$\sim$2,  consistently  with the results by \citet{calzetti00}. \\
Logarithm of the stellar masses ranges between 9.8 and 10.5 indicating that 
we are observing the most massive star-forming galaxies in the cluster, 
missing the fainter population of less massive and dwarf galaxies.  

\begin{figure}
\begin{center} 
\includegraphics[width=8.2truecm]{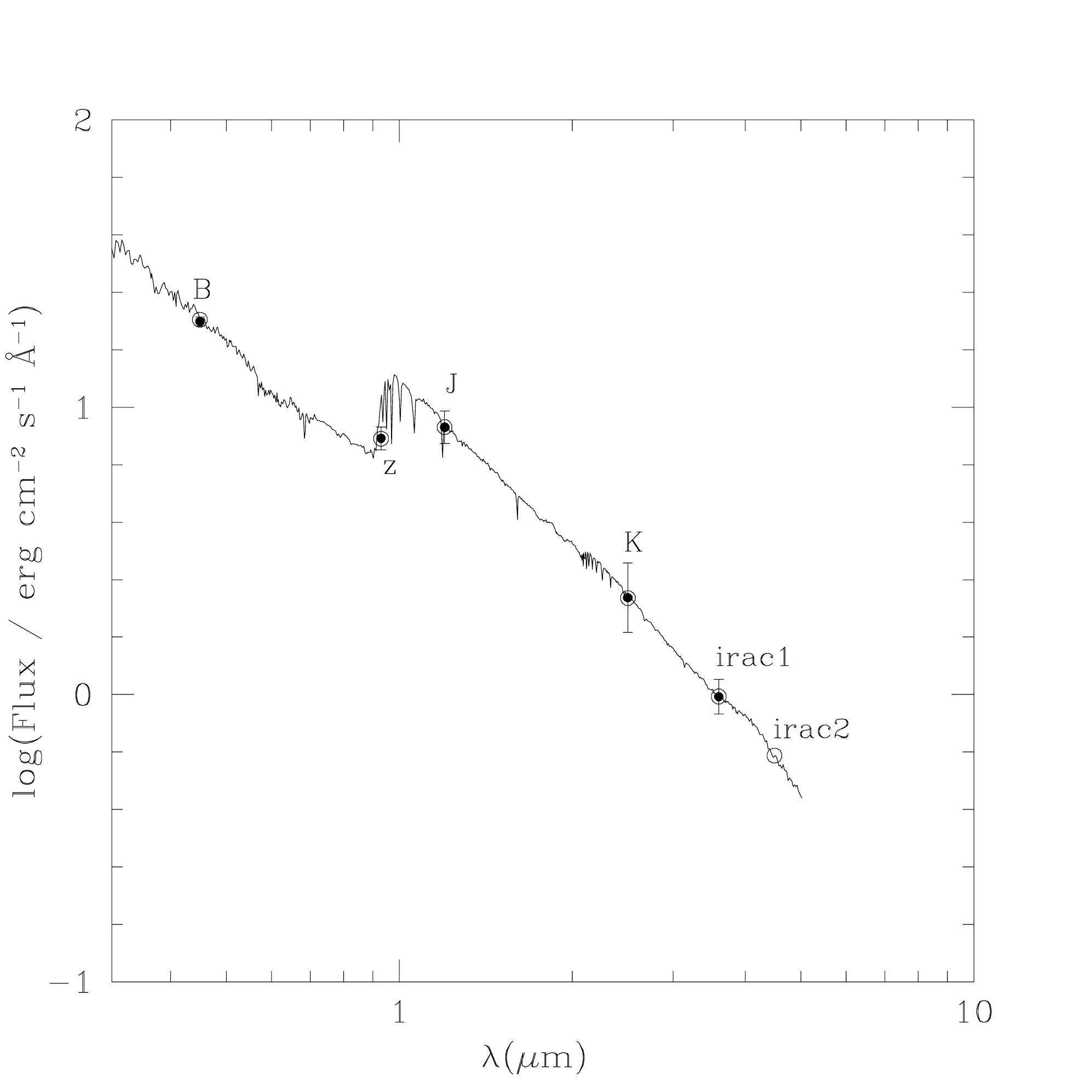} 
\caption{An example of our SED fitting for the galaxy MSC~2: the continuous line is the best fitting, the filled circles are the measured fluxes, and the empty circles  are the fluxes expected on the basis of the best fitting.}
\end{center}
\label{fig_sed}
\end{figure}
\subsection[]{Metallicity determination}
\label{sec_met}
We used  N2$=$[NII]/H$\alpha$  ratio to infer the metallicity of the galaxies using 
the calibrations obtained by \citet{nagao06} and \citet{maiolino08}. 
 The wavelength separation between the two lines is very small and thus this ratio is not affected by the  extinction effects. 
The uncertainties of the gas-phase metallicities, of the order of $0.1$-$0.2$ dex
result from both the error on the calibration itself and the error on the estimation
of [NII] emission line (that can reach $50$-$70$\%, while typical error on the H$\alpha$ fluxes are $10$\%). 
Even if N2 is commonly adopted as a metallicity indicator, it is necessary to remember the intrinsic limit of the metallicity measurement with N2 which depends not only on O/H, but also 
on N/O. 
The relation between N2 and 12$+\log$(O/H) is based on the observations of star-forming  galaxies or single HII regions in the Local Universe. 
The relation is not linear:  at low oxygen abundances the primary production of nitrogen dominates (N/O is constant),  at high oxygen abundance  the secondary production of nitrogen dominates
(N/O increases with 12$+\log$(O/H)).
The observed relation has however a large scatter since N/O ratio is also a clock of the last 
major episode of star formation \citep[e.g.,][]{vanzee98}. 
Low values of N/O might not only imply low metallicity, but they might be the signature of  a very recent burst of star formation, while high values of 
N/O might also imply galaxies with a long quiescent period. 
This effect is more dominant in dwarf galaxies whose SF is characterized by bursty episodes, while massive galaxies have a more continuous star formation history
\citep[see also][]{perez08,zahid12,maiolino12}.

\subsection[]{The Star Formation Rate}
\label{sec_sfr}
We have derived the SFR from the H$\alpha$ flux, which provides a direct probe of the young massive stellar population. 
We have used the relationship given by \citet{kennicutt98}, and 
scaled down the results by a factor of 1.7 \citep{pozzetti07} to convert them to the \citet{chabrier03} IMF, as described by \citet{cresci12}.  
The conversion factor is based on the H$\alpha$ flux corrected for dust extinction. 
Thus  a good knowledge of the amount of extinction which affects the emission lines is required.
Since we were not able to measure the Balmer ratio H$\alpha$/H$\beta$, from which the extinction can be derived, 
we have used the extinction A$_{V}$ derived from the SED fitting described in Sec.\ref{mass}.   
The errors on the SFR take into account the error on the H$\alpha$ flux ($\sim$10\%) and the formal error in the extinction (0.2 mag). 
Consistently with what discussed in Sec.\ref{mass}, we have also computed the SFRs using  $A_{V}$ value multiplied by a factor 2, which seem to 
be more appropriate for the gaseous component \citep{forster09}.  
In Tab.\ref{tab_fmr} we present the SFR values obtained with $A_{V}$ from SED fitting, while in Fig.\ref{fmr} (right panel) we 
show both values of $\mu$=$\log$M-$\alpha$ $\log$SFR, obtained with the SFR corrected with nebular (empty circles) and stellar (filled circles) $A_V$.

\label{tab_mass}
\begin{table*}
\caption{Galaxy properties}
\centering
\scriptsize
\begin{tabular}{cccccccllcc}
\hline
\multicolumn{4}{c}{}\\
 id  &  B & z & J & K & IRAC3.6 & IRAC4.5&   log(\mstar)             & 12+log($\frac{O}{H}$)          &  log(SFR)   & A$_{V}$       \\
\multicolumn{4}{c}{}\\
\hline
ser.1 	& 23.06$\pm0.06$ & 23.11$\pm0.13$  & 22.16$\pm0.13$ & 21.02$\pm0.14$  & $-$ 			& 21.12$\pm0.14$ 	&10.4$\pm$0.15	         & 8.82$\pm$0.12 	& 0.7$\pm$0.2  	&0.2$\pm$0.2\\
sBzK.7208 	& 24.93$\pm0.07$ & 23.46$\pm0.14$ & 22.00$\pm0.22$ & $-$       			& 21.29$\pm0.06$ 	& 21.33$\pm0.11$ 	&10.0$\pm$0.2 	         & 8.70$\pm$0.12  	& 1.4$\pm$0.2	&1.9$\pm$0.2 \\
ser.3 	& 25.29$\pm0.11$ & 23.18$\pm0.14$ & 21.50$\pm0.15$ & $ -$ 			& 21.27$\pm0.05$ 	& 21.77$\pm0.15$ 	&10.5$\pm$0.15 	         & 8.96$\pm$0.14 	& 0.9$\pm$0.2	&0.0$\pm$0.2\\
MSC1     & 23.57$\pm0.04$ & 21.71$\pm0.10$ & 21.68$\pm0.14$ & 22.87$\pm0.30$ 	& 22.54$\pm0.15$ 	& $-$   			& 9.8$\pm$0.2 		& 8.62$\pm$0.14  	& 1.5$\pm$0.2	& 0.0$\pm$0.2 	\\
MSC2 	& 23.58$\pm0.04$ & 22.72$\pm0.11$ & 22.66$\pm0.16$ & $-$ 			& $- $ 			& $-$ 			& 9.8$\pm$0.2	         & $<$8.92$^{a}$  	  	& 0.03$\pm$0.2	&1.1$\pm$0.2 	\\
\multicolumn{4}{c}{}\\
\hline
\end{tabular}
(a) upper limit on [NII] measurement

\label{tab_fmr}
\end{table*}

\section{The mass-metallicity and the fundamental metallicity relations}
\label{sec_fmr}

In the Local Universe, there is not an unanimous agreement on the effect of cluster membership in the chemical evolution of galaxies. 
\citet{ellison09} presented data for 1318 galaxies in local clusters, obtaining their stellar mass and gas-phase abundances. 
By comparing the MZR of the cluster galaxies with a control sample of galaxies matched in
mass and  redshift, they  found that cluster galaxies  have, on average, at a given mass higher metallicities by up to 0.04 dex.
However they found that this effect is not related simply to cluster membership, nor to cluster properties. 
They attributed it to local scale processes, such
as the presence of a close companion or several near neighbours
that lead to an enhanced metallicity.
On the other hand,  \citet{petro12}  from the detailed analysis of dwarf galaxies four nearby clusters, have found  that
the enhancement of the gas-phase metallicity is sensitive to the cluster mass, and appears to be more prominent in the inner regions of massive clusters. 
They claim that this evidence points towards a possible connection of the chemical
enrichment of cluster galaxies with their ICM properties.
To clarify these aspects we are planning to present an analysis of the effect of the environment on the FMR in galaxies in nearby clusters 
in a forthcoming  paper. 

In the high-redshift Universe, the situation is even less defined due  to the limited number of works dedicated 
to this issue. 
Most of the works dedicated to clusters at z$>$1 focussed their attention 
on the  spatial distribution of star-forming and non-star-forming galaxies. 
\citet{hayashi11}  analyzed a sample of star-forming galaxies at z$=$1.46 and they found 
that the galaxies located close to the cluster core are experiencing high star-forming activity
comparable to those in other lower-density regions. 
\citet{fassenberd09} also found an on-going
starburst activity in a cluster at z$=$1.56, and from a comparison with literature data they concluded that most
of the clusters at z$\sim$1.5 are likely to hold active star formation in
the core regions.
On the contrary, in other clusters with slightly lower redshifts the star formation activity has been found
significantly weaker than that in the surrounding regions \citep[at z$=$0.8 and z$=$1.39,][]{Lidman08,koyama10,bauer11}.

To our knowledge, only  \citet{hayashi11}  have intended to investigate
the metallicity in a cluster at $z > 1$ observing  several star-forming galaxies in the cluster XMMXCSJ2215.9-1738 at z$=$1.46.
Their results suggest that the metallicity 
in that cluster is  not (yet) strongly dependent on environment and 
that the cluster galaxies  are located on a MZR, 
similarly  to that of the star-forming galaxies in the field at z$\sim$2.  
However, due to non-photometric weather conditions, \citet{hayashi11} were not able to calibrate their H$\alpha$ flux and thus 
a direct comparison with our SFRs is not feasible. 

\begin{figure} 
   \centering
   \includegraphics[width=10.2truecm]{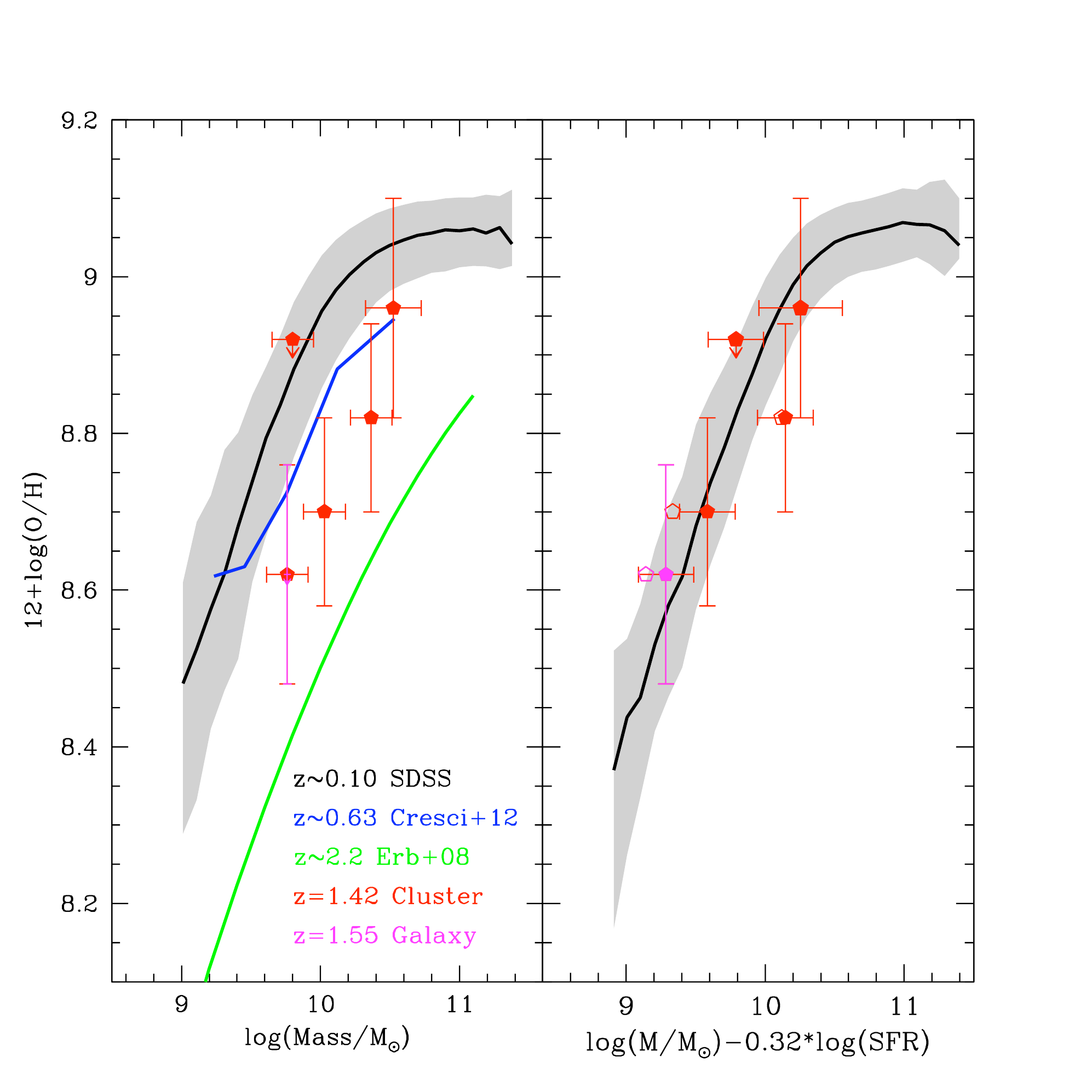} 
 \caption{The FMR in the cluster 7C 1756+6520: left panel, the MZR as defined by the SDSS galaxies at z$\sim$0.1 (black curves, and shaded area), at z$\sim$0.63 (blue curve), 
 and at z$\sim$2.2 (green curve); right panel,  the projection of the FMR 
 in the plane $\mu$=$\log$M-$\alpha$ $\log$SFR, where $\alpha=0.32$. In both panels red filled circles are the galaxies observed in  the cluster 7C 1756+6520, while the magenta filled circle is the galaxy
 projected on the cluster area with a redshift z$=$1.55 (SFR computed with the extinction from SED fitting). 
 The empty circles  show the $\mu$ values obtained with the SFR with the nebular extinction.} 
   \label{fmr}
\end{figure}

With our observations of the cluster 7C 1756+6520 we are adding another piece of information to the
view of the evolution of galaxies within a cluster by having the
complete information of the stellar masses, SFR, and metallicity. 
\citet{mannucci10} have shown that it exists a combination of the stellar mass and SFR which better correlates with the metallicity. 
In Fig.\ref{fmr} we show in the left panel the {\em classical} MZR and its evolution with redshift and in the right panel the projection of the FMR in the plane $\log$M-$\alpha$ $\log$SFR, where $\alpha=0.32$. 
The shaded area represent the local SDSS galaxies at z$\sim$0.1 (see \citet{mannucci10}). 
We note immediately an evolution of the MZR relation with the redshift: the galaxies of the cluster at z$\sim$1.4 do not follow to local MZR, having at a given 
mass a lower metallicity. The MZR of the cluster is indeed in between the MZR derived by \citet{cresci12} at z$\sim$0.63 and by \citet{erb08} at z$\sim$2.2. 

\begin{figure} 
   \centering
   \includegraphics[width=9.2truecm]{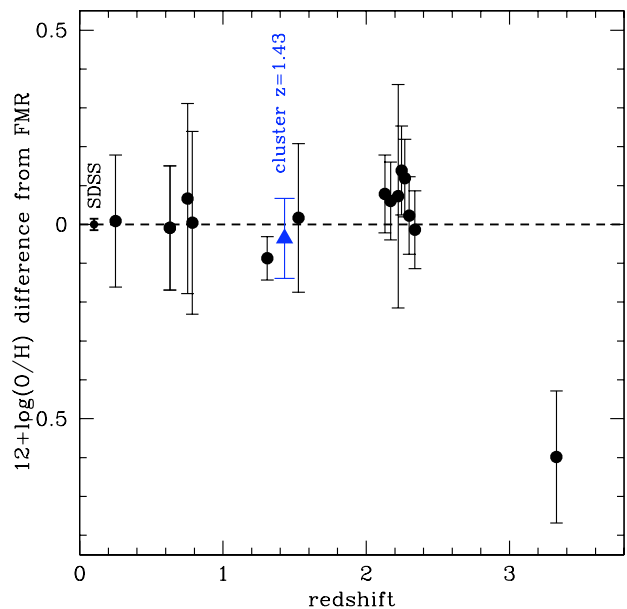} 
 \caption{The metallicity difference from the FMR (defined with SDSS galaxies at z$\sim$0.1) for galaxies at different redshifts. 
For field galaxies the results of \citet{mannucci10} is that all  galaxies up to z$=$2.5 are consistent with no evolution of the FMR. The filled circles are from \citet{mannucci10}, the filled triangle 
is the result of the present wotk. } 
   \label{fmr1}
\end{figure}

When we apply to the MZR the correction for the SFR, we found that the galaxies of the cluster lie in the locus of the Local Universe galaxies, the so-called FMR which is valid 
for field galaxies up to z$\sim$2.5. 
Note that in the plot we have also included the galaxy MSC~1 (in magenta), which has a redshift of 1.55, thus clearly outside the cluster. 
The galaxies belonging to the two sub-groups of the cluster  7C 1756+6520 (filled red circles) are in good agreement with the FMR derived from isolated galaxies. 
This is also true when a the nebular extinction (two times the extinction derived from SED fitting) is applied. This affect marginally the value of $\mu$ for 
the galaxies with a higher extinction (empty circles in Fig.\ref{fmr}).\\
Is this absence of deviation expected? or is it telling us something new about the evolution of galaxies in cluster in the early phases of their evolution?
In Fig.\ref{fmr1} we show the metallicity difference from the FMR (defined with SDSS galaxies at z$\sim$0.1) for galaxies at different redshifts. 
For field galaxies the result of \citet{mannucci10} is that all  galaxies up to z$=$2.5 are consistent with no evolution of the FMR.
Our observations, even if affected by a small statistics, suggest that star-forming galaxies in clusters at  z$\sim$1.4 follow 
the FMR in the same way as field galaxies\footnote{The average has been computed without including the upper limit value}.  As shown in Fig.~4 the average difference of our galaxy  metallicity with the FMR is
very small. 
Our results indicate  that,  in high redshift clusters,  the chemical evolution is driven by the intrinsic {\em nature} of the galaxy evolution, 
and that external perturbations are not observable in terms of location of galaxies in the FMR. However, the uncertainties  on our metallicity measurements prevent us 
to estimate differences of the order of 0.04~dex, as observed by \citet{ellison09} between cluster and field galaxies.  
A comparison with the FMR in the Local Universe clusters is necessary, and will be the argument of a forthcoming paper. 

A possible explanation to this lack of effects due to cluster membership is the large extension of the 
structure around the radio galaxy 7C 1756+6520 and in high-z clusters in general. 
The cosmological simulation of \citet{poggianti10} found that the  number of galaxies in cluster per unit volume is higher by approximately a factor of 4.7 at z= 1.5, compared to z= 0.
They also found that the most massive and the least massive clusters or groups are on average equally dense, at a given redshift.
However, these considerations  apply to virialized regions, and probably are not valid in the case of large scale structure as that around 7C 1756+6520. 
Often over-densities at the highest redshifts  have a filamentary nature and extend beyond $\sim $2 Mpc \citep{croft05,zappacosta02,mannucci07}. 
As described by G10, 7C 1756+6520 is part of a large scale galaxy structure composed of (at least) two main groups - a galaxy cluster centered on the radio galaxy at z=1.4156  and a compact group at $z\sim1.437$, $1.5\hbox{$^\prime$ }$ east of the radio galaxy. 
In this filamentary structure  the influence of galaxy  interaction or of the  ICM has not  modified  the galaxy evolution enough to 
change their position in the FMR plane.   
Observations of the X-ray emission of the ICM of this cluster would be extremely useful to study its dynamical state. 

%
\section[]{The bright AGN 1317}
\label{sec_agn}
\begin{figure*} 
   \hbox{
   \includegraphics[width=0.5\linewidth]{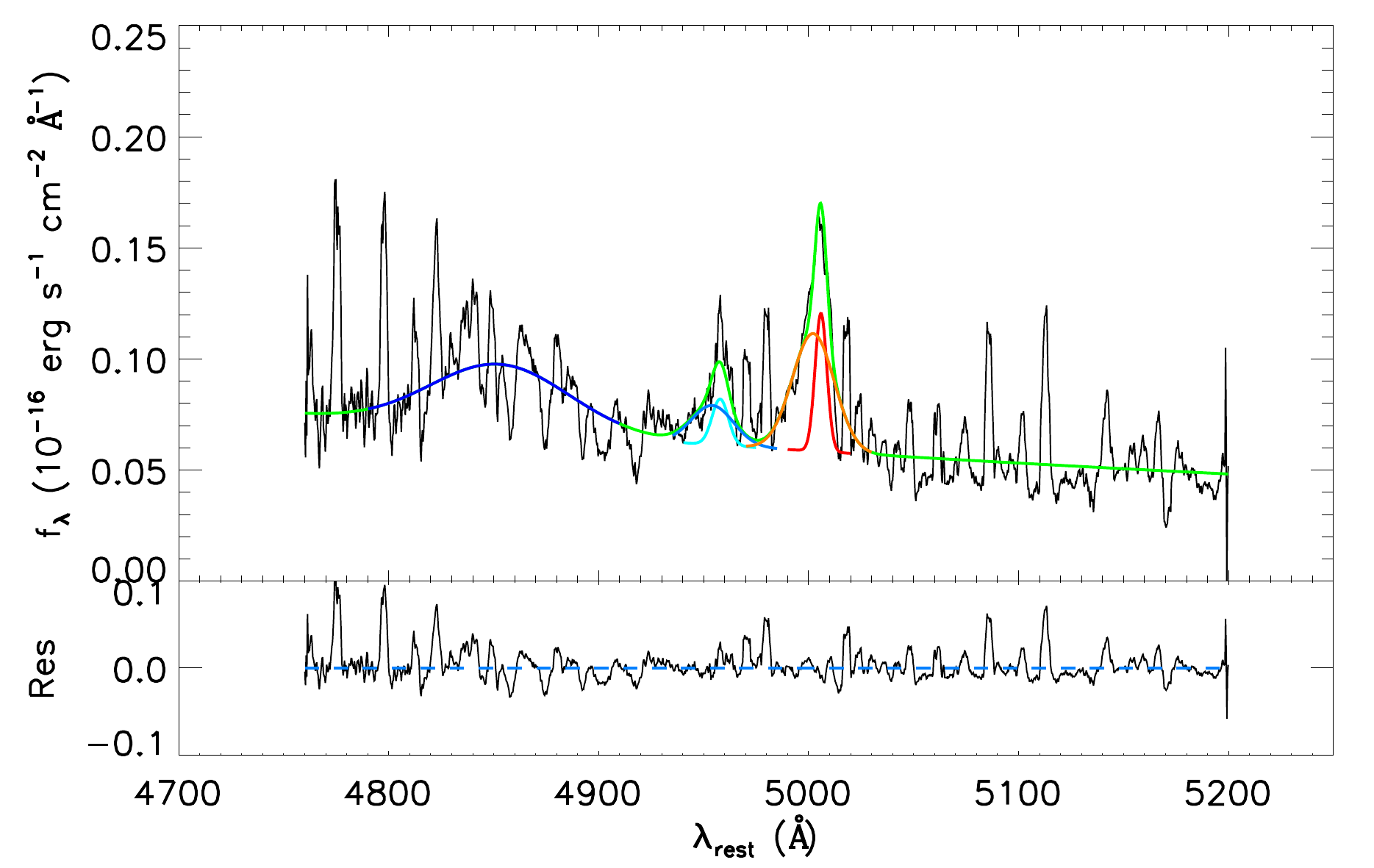} 
   \includegraphics[width=0.5\linewidth]{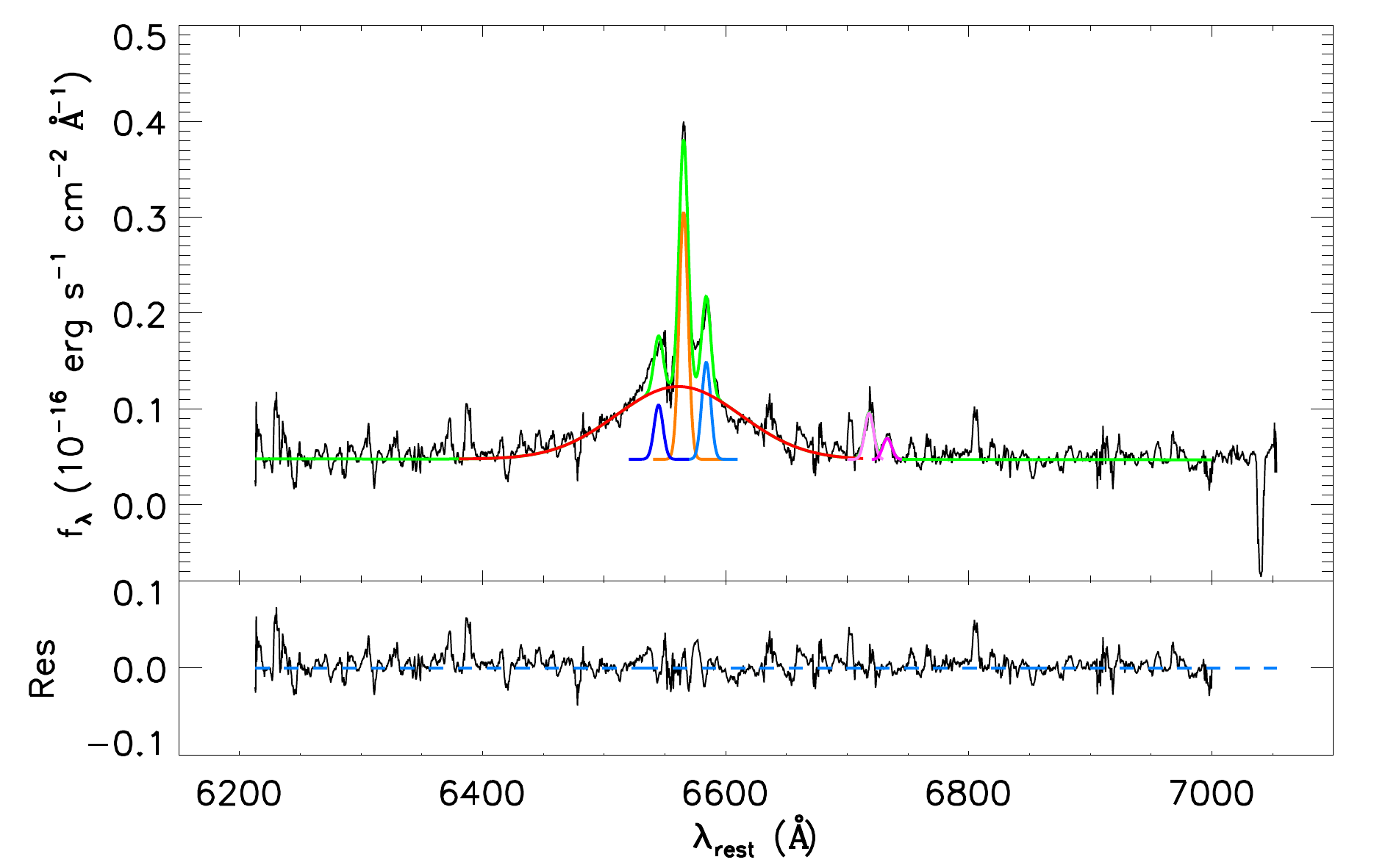}
   }
 \caption{AGN.~1317 rest frame spectrum. \textit{Left}: J-band data, normalized to match the H-band continuum. \textit{Right}: H-band data.
 In both panels, the green line represents the global fit to the data with residuals shown in the bottom board. The reduced $\chi^2$ are 1.7 for 
 the J-band and 1.3 for the H-band fitting.
 Emission features are fitted by Gaussian shapes. In the \textit{left} panel, dark blue is for the H$\beta$ ($4861\mbox{\AA}$), 
 sky blue and cyan for the [OIII]$\lambda 4959$ (narrow and broad components, respectively), orange and red for the 
 the [OIII]$\lambda 5007$ (narrow and broad components, respectively). 
 In the \textit{right} panel, red and orange for the H$\alpha$ ($6563\mbox{\AA}$, 
 broad and narrow components, respectively), blue and sky blue for the [NII] lines ($6548\mbox{\AA}$ and $6584\mbox{\AA}$,
respectively), while light and dark pink are respectively for the [SII]$\lambda 6716\mbox{\AA}, \lambda 6731\mbox{\AA}$.} 
   \label{agn}
\end{figure*}
Seven  AGNs, including  the central radio galaxy, have  been spectroscopically confirmed 
in close proximity both spatially and in redshift space of the cluster. 
Three of them are found within $1.5\hbox{$^\prime$ }$ of the radio galaxy, including AGN.1317. 
We present the spectroscopic observations of AGN.1317 both in J and H bands. 
As mentioned in Section~\ref{sec_obs}, it was not possible to perform a proper flux calibration of the J-band data. 
To obtain a rough calibration, we re-scaled the J-band data to match the H-band continuum by a simple visual comparison of the two spectral bands. 
The spectrum of AGN.1317 (Fig.~\ref{agn}) shows clear broad features, as \hb (FWHM = $4800\pm500$~km/s) and \ha  
(FWHM = $5520\pm250$~km/s), associated with the broad line region (BLR), 
and forbidden lines primarily associated with the AGN narrow line region (NLR): the [OIII], [NII] and [SII] doublets. 

The [OIII]$\lambda\lambda 4959,5007$ lines show a clear asymmetric profile with prominent blue-shifted wings. This is a characteristic 
signature of outflow. In fact, as observed in local AGN \citep{muller11}, 
the net outflow of the NLR clouds through a dusty region (of the host galaxy) implies 
that the emission from the clouds on the far, red-shifted side is suppressed relative to the emission on the near, 
blue-shifted side. 

The [OIII]$\lambda\lambda 4959,5007$ lines are well fitted by two Gaussians (narrow and blue-shifted components) as seen in Fig~\ref{agn}. 
The gaussians of the doublet of narrow lines are forced to have the theoretical intensity ratio 
(i.e. [OIII]$\lambda 4959$/[OIII]$\lambda 5007$=1/3), the same FWHMs, and their known rest-frame wavelengths. 
The blue-shifted components are treated in a similar way,  but we allow  the central wavelengths  to vary. 
This yields a FWHM$\sim 550$~km/s for the narrow components, nicely consistent with typical NLR values. 
The asymmetric wings are reproduced by broader Gaussians (FWHM$\sim 1420$~km/s) blue-shifted by 430$\pm$180~km/s. 
These FWHMs are much broader 
than usually found in the NLR of local AGNs \citep{muller11} and are rather consistent with what observed 
in powerful quasars at higher redshift \citep{cano12}. 
The [OIII] doublet  primarily traces the gas in the NLR ionized by the AGN, 
thus the most probable explanation is that the strong outflow might be  driven by the AGN radiation pressure itself. 
The origin of the wind from the AGN is also supported by the huge velocities (the out-flowing gas reaches an excess of 1000~km/s) 
which cannot be explained by supernovae-driven outflows \citep{nath09}.
This is the first time that such  powerful AGN-driven winds are detected in source located at the edges of a high redshift cluster 
(the projected distance between the central radio galaxy and AGN.~1317 being $\sim 780$~kpc).
Indeed, at z$> 1$ AGN outflows traced by [OIII] emission have been detected only in isolated sources so far 
\citep{nesvadba08, nesvadba11, alexander10}. 
However, to draw some conclusion about the significance of this outflow (i.e. the out-flowing mass, at which rate, whether the outflow can 
affect the cluster environment, etc.), it is necessary a follow-up with calibrated J-band data. 

\section{Conclusions}
\label{sec_discu}

We have presented new infrared spectroscopic observations with LUCIFER@LBT of a sample of star-forming galaxies in the two main sub-groups of the cluster 
associated with the radio galaxy 7C1756+6520.
With available photometric data and with our spectroscopic information, we have derived  their stellar mass, metallicity and SFR. 
We have then located them in the MZ plane and in the FMR. The galaxies in the z$\sim1.4$ cluster are perfectly consistent with the FMR, 
suggesting that the effect of the environment is not dominant in the early phases of their evolution. 
Finally, we have detected in the [OIII] doublet a strong gas outflow reaching velocities $>1000$~km/s that is 
possibly driven by the AGN radiation pressure. This source, AGN.~1317, is located 
at the edges of the cluster and further observations are required to study the effect of such strong wind on both the host and cluster environments.

\section*{Acknowledgments}
L. Magrini and E. Sani acknowledge financial support from ASI under Grant I/009/10/0/.

\bsp

\label{lastpage}


\begin{thebibliography}{}
\bibitem[Ageorges et al.(2010)]{ageorges10} Ageorges, N., Seifert, W., Jutte, M., et al., 2010, SPIE, 7735, 53 
\bibitem[Alexander et al. (2010)]{alexander10} Alexander D.~M., Swinbank A.~M., Smail I., McDermid R., Nesvadba N.~P.~H., 2010, MNRAS, 402, 2211 
\bibitem[Arnaud(2009)]{arnaud09} Arnaud, M., 2009, A\&A, 500, 103
\bibitem[Bauer et al.(2011)]{bauer11} Bauer, A.~E.,  Gr{\"u}tzbauch, R., J{\o}rgensen, I., Varela, J., \& Bergmann, M.\ 2011, MNRAS, 411, 2009 
\bibitem[Blakeslee et al.(2003)]{blakeslee03} Blakeslee, J.~P.,  Franx, M., Postman, M., et al.\ 2003, ApJL, 596, L143 
\bibitem[Bolzonella et al.(2010)]{bolzonella10}  Bolzonella, M., Kova, K.,  Pozzetti, L.,  et al., 2010, A\&A, 524, 76
\bibitem[Bolzonella et al.(2000)]{bolzonella00} Bolzonella, M., Miralles, J.-M., \& Pell{\'o}, R.\ 2000, \aap, 363, 476 
\bibitem[Bruzual \& Charlot(2003)]{bruzual03} Bruzual, G., \& Charlot, S., 2003, MNRAS, 344, 1000
\bibitem[Calzetti et al.(2000)]{calzetti00}  Calzetti, D., Armus, L., Bohlin, R. C.,  et al., 2000, ApJ, 533, 682
\bibitem[Campisi et al.(2011)]{campisi11} Campisi, M. A., Tapparello, C., Salvaterra, R., et al., 2011, MNRAS, 1437
\bibitem[Cano-D{\'{\i}}az et al. (2012)]{cano12} Cano-D{\'{\i}}az M., Maiolino R., Marconi A., Netzer H., Shemmer O., Cresci G., 2012, A\&A, 537, L8
\bibitem[Cresci et al(2012)]{cresci12} Cresci, G.,  Mannucci, F., Sommariva, V., et al, 2012, MNRAS, tmp.2212 
\bibitem[Croft et al.(2005)]{croft05} Croft, S.,  Kurk, J.,  van Breugel, W., et al., 2005, AJ, 130, 867 
\bibitem[Chabrier(2003)]{chabrier03} Chabrier, G., 2003, ApJ, 586, 133
\bibitem[Chung et al.(2009)]{chung09} Chung, A., van Gorkom, J.~H., Kenney, J.~D.~P., Crowl, H., \& Vollmer, B.\ 2009, \aj, 138, 1741 
\bibitem[Daddi et al.(2004)]{daddi04} Daddi, E., Cimatti, A.,  Renzini, A., et al., 2004, ApJ, 617, 746 
\bibitem[Dav\'e et al.(2011)]{dave11} Dav\'e, R., Finlator, K., Oppenheimer, B. D., 2011, MNRAS, 416, 135
\bibitem[Davies(2007)]{davies07} Davies, R.~I.\ 2007, \mnras, 375, 1099 
\bibitem[Dayal et al.(2012)]{dayal12} Dayal, P., Ferrara, A., \& Dunlop, J.~S.\ 2012, arXiv:1202.4770 
\bibitem[di Serego Alighieri et 
al.(2005)]{diserego05} di Serego Alighieri, S., Vernet, J., Cimatti, A., et al.\ 2005, \aap, 442, 125 
\bibitem[Dressler et al.(1980)]{dressler80} Dressler, A., 1980, ApJ, 236, 351
\bibitem[Ellison et al.(2009)]{ellison09} Ellison, S. L.,  Simard, L.,  Cowan, N. B., et al., 2009, MNRAS, 397, 467
\bibitem[Erb et al.(2006)]{erb06} Erb, D. K., Steidel, C. C., Shapley, A. E.,  et al., 2006, ApJ, 646, 107
\bibitem[Erb(2008)]{erb08} Erb, D.~K.\ 2008, \apj, 674, 151 
\bibitem[Eisenhardt et al.(2008)]{eisenhardt08} Eisenhardt, P.~R.~M., Brodwin, M., Gonzalez, A.~H., et al.\ 2008, \apj, 684, 905 
\bibitem[Fassbender et al.(2009)]{fassenberd09} Fassbender, R.,  Nastasi, A.,  Boehringer, H., et al., 2011, A\&A, 527, 10 
\bibitem[Finlator \& Dav\'e(2008)]{finlator08} Finlator, K., \& Dav\'e, R., 2008, MNRAS, 385, 2181
\bibitem[Finn et al.(2005)]{finn05} Finn, R. A.,  Zaritsky, D., McCarthy, D. W., Jr., et al., 2005, ApJ, 630, 206  
\bibitem[F{\"o}rster Schreiber et al.(2009)]{forster09} F{\"o}rster Schreiber, N.~M., Genzel, R., Bouch{\'e}, N., et al.\ 2009,  \apj, 706, 1364 
\bibitem[Galametz et al.(2009)]{g09} Galametz, A., De Breuck, C., Vernet, J., et al. 2009, A\&A, 507, 131 (G09)
\bibitem[Galametz et al.(2010)]{g10} Galametz, A., Stern, D., Stanford, S.~A., et al.\ 2010, \aap, 516, A101 (G10)
\bibitem[Galametz et al.(2010B)]{g10b} Galametz, A., Vernet, J., De Breuck, C., et al.\ 2010, \aap, 522, A58 
\bibitem[Gobat et al.(2011)]{gobat11} Gobat, R., Daddi, E., Onodera, M., et al.\ 2011, \aap, 526, A133 
\bibitem[Hayashi et al.(2011)]{hayashi11} Hayashi, M., Kodama, T.,  Koyama, Y., et al., 2011,MNRAS, 415, 2670 
\bibitem[Hatch et al.(2011)]{hatch11} Hatch, N. A., De Breuck, C., Galametz, A., et al.\ 2011, MNRAS, 410, 1537 
\bibitem[Hill et al.(2006)]{hill06} Hill, J. M.,  Green, R. F.,  \& Slagle, J. H., 2006, SPIE, 6267, 31
\bibitem[Hilton et al.(2007)]{hilton07} Hilton, M., Collins, C. A., Stanford, S. A., et al. 2007, ApJ, 670, 1000 
\bibitem[Hilton et al.(2009)]{hilton09} Hilton, M.,  Stanford, S. A.,  Stott, J. P., et al., 2009, 697, 436 
\bibitem[Horne(1986)]{horne86} Horne, K.\ 1986, \pasp, 98, 609 
\bibitem[Kennicutt(1998)]{kennicutt98} Kennicutt, R.~C., Jr.\ 1998, \araa, 36, 189 
\bibitem[Koyama et al.(2010)]{koyama10}  Koyama, Y.,  Kodama, T.,  Shimasaku, K., et al., 2010, MNRAS, 403, 1611
\bibitem[Kuiper et al.(2011)]{kuiper11} Kuiper, E., Hatch, N.~A., Venemans, B.~P., et al.\ 2011, MNRAS, 417, 1088
\bibitem[Lequeux et al.(1979)]{lequeux79}  Lequeux, J., Peimbert, M.,  Rayo, J. F., et al., 1979, A\&A, 80, 155 
\bibitem[Lidman et al.(2008)]{Lidman08} Lidman, C., Rosati, P., Tanaka, M., et al. 2008, A\&A, 489, 981 
\bibitem[Maiolino et al.(2008)]{maiolino08}  Maiolino, R., Nagao, T., Grazian, A., et al., 2008, A\&A, 488, 463  
\bibitem[Maiolino et al.(2012)]{maiolino12}  Maiolino, R. et al., 2012, in prep.
\bibitem[Mannucci et al.(2007)]{mannucci07} Mannucci, F., Bonnoli, G., Zappacosta, L., Maiolino, R., \& Pedani, M.\ 2007, \aap, 468, 807 
\bibitem[Mannucci et al.(2009)]{mannucci09}  Mannucci, F., Cresci, G., Maiolino, R.,  et al., 2009, MNRAS, 398, 1915
\bibitem[Mannucci et al.(2010)]{mannucci10}  Mannucci, F., Cresci, G., Maiolino, R., et al., 2010, MNRAS, 408, 2115
\bibitem[Mannucci et al.(2011)]{mannucci11}  Mannucci, F., Salvaterra, R., \& Campisi, M. A., 2011, MNRAS, 414, 1263
\bibitem[M{\"u}ller-S{\'a}nchez et al. (2011)]{muller11} M{\"u}ller-S{\'a}nchez F., Prieto M.~A., Hicks E.~K.~S., Vives-Arias H., Davies R.~I., Malkan M., Tacconi L.~J., Genzel R., 2011, ApJ, 739, 69 
\bibitem[Nagao et al.(2006)]{nagao06}  Nagao, T., Maiolino, R., \& Marconi, A., 2006, A\&A, 459, 85
\bibitem[Nakata et al.(2005)]{nakata05}  Nakata, F.,  Kodama, T.,  Shimasaku, K., et al., 2005, MNRAS, 357, 1357 
\bibitem[Nath \& Silk (2009)]{nath09} Nath B.~B., Silk J., 2009, MNRAS, 396, L90 
\bibitem[Nesvadba et al. (2008)]{nesvadba08} Nesvadba N.~P.~H., Lehnert M.~D., De Breuck C., Gilbert A.~M., van Breugel W., 2008, A\&A, 491, 407 
\bibitem[Nesvadba et al. (2011)]{nesvadba11} Nesvadba N.~P.~H., Polletta M., Lehnert M.~D., Bergeron J., De Breuck C., Lagache G., Omont A., 2011, MNRAS, 415, 2359 
\bibitem[Papovich(2008)]{papovich08} Papovich, C.\ 2008, \apj, 676, 206 
\bibitem[Papovich et al.(2010)]{papovich10} Papovich, C., Momcheva, I., Willmer, C.~N.~A., et al.\ 2010, \apj, 716, 1503 
\bibitem[P{\'e}rez-Montero \& Contini(2009)]{perez08} P{\'e}rez-Montero E., Contini T., 2009, MNRAS, 398, 949
\bibitem[Petropoulou et al.(2012)]{petro12} Petropoulou, V., Vilchez, J.~M., \& Iglesias-Paramo, J.\ 2012, arXiv:1202.4164 
\bibitem[Poggianti et al.(2006)]{poggianti06} Poggianti, B. M.,  von der Linden, A.,  De Lucia, G., et al., 2006, ApJ, 642, 188 
\bibitem[Poggianti et al.(2010)]{poggianti10} Poggianti, B. M., De Lucia, G.,  Varela, J., et al., 2010, MNRAS, 405, 995 
\bibitem[Pozzetti et al.(2007)] {pozzetti07} Pozzetti, L., Bolzonella, M., Lamareille, F., et al., 2007, A\&A, 474, 443
\bibitem[Pozzetti et al.(2010)]{pozzetti10} Pozzetti, L., Bolzonella, M., Zucca, E.,  et al., 2010, A\&A, 523, 13
\bibitem[Rosati et al.(2004)]{rosati04} Rosati, P., Tozzi, P.,  Ettori, S., et al., 2004, AJ, 127, 230 
\bibitem[Rosati et al.(2009)]{rosati09} Rosati, P., Tozzi, P.,  Gobat, R., et al., 2009, A\&A, 508, 583 
\bibitem[Scodeggio et al.(2005)]{scodeggio05} Scodeggio, M., Franzetti, P., Garilli, B., et al.\ 2005, \pasp, 117, 1284 
\bibitem[Seifert et al.(2010)]{seifert10} Seifert, W., Ageorges, N., Lehmitz, M., et al., SPIE, 7735, 256
\bibitem[Sommariva et al.(2011)]{sommariva12} Sommariva, V., Mannucci, F., Cresci, G., et al.\ 2011, \aap, 539, A136 
\bibitem[Stanford et al.(2005)]{stanford05} Stanford, S. A.,  Eisenhardt, P. R.,  Brodwin, M., et al., 2005, ApJ, 634, 129 
\bibitem[Stanford et al.(2006)]{stanford06} Stanford, S. A., Romer, A. K.,  Sabirli, K., et al., 2006, ApJ, 646, 13
\bibitem[Tanaka et al.(2009)]{tanaka09}Tanaka, M., Lidman, C.,  Bower, R. G., et al., 2009, A\&A, 507, 671
\bibitem[Tanaka et al.(2010)]{tanaka10} Tanaka, M.,  De Breuck, C., Venemans, B., et al, 2010, A\&A, 518, 18
\bibitem[Tremonti et al.(2004)]{tremonti04} Tremonti, C. A., Heckman, T. M., Kauffmann, G., et al. 2004, ApJ, 613, 898  
\bibitem[van Zee et al.(1998)]{vanzee98} van Zee, L., Salzer, J.~J., \& Haynes, M.~P.\ 1998, ApJL, 497, L1 
\bibitem[Venemans et al.(2007)]{venemans07} Venemans, B.~P., R{\"o}ttgering, H.~J.~A., Miley, G.~K., et al.\ 2007, \aap, 461, 823 
\bibitem[Wilson et al.(2009)]{wilson09} Wilson, G., Muzzin, A., Yee, H.~K.~C., et al.\ 2009, \apj, 698, 1943 
\bibitem[Zahid et al.(2012)]{zahid12} Zahid, H.~J., Bresolin, F., Kewley, L.~J., Coil, A.~L., \& Dav{\'e}, R.\ 2012, arXiv:1203.0558 
\bibitem[Zappacosta et al.(2002)]{zappacosta02} Zappacosta, L., Mannucci, F., Maiolino, R., et al.\ 2002, \aap, 394, 7 


\end{thebibliography}
\end{document}